%% file: main.tex
\documentclass[11pt,a4paper]{article}

\usepackage{graphicx}
\usepackage{multirow}
\usepackage{amsmath,amssymb,amsfonts}
\usepackage{amsthm}
\usepackage{mathrsfs}
\usepackage[title]{appendix}
\usepackage{xcolor}
\usepackage{textcomp}
\usepackage{manyfoot}
\usepackage{booktabs}
\usepackage{algorithm}
\usepackage{algorithmicx}
\usepackage{algpseudocode}
\usepackage{listings}

\usepackage{siunitx}
\usepackage{physics}
\usepackage{array}
\usepackage{arydshln}
\usepackage{caption}
\usepackage{subcaption}
\usepackage{environ}
\usepackage{csquotes}
\usepackage{authblk}
\usepackage{geometry}
\usepackage{hyperref}

\AtBeginDocument{\RenewCommandCopy\qty\SI}
\ExplSyntaxOn
\msg_redirect_name:nnn{siunitx}{physics-pkg}{none}
\ExplSyntaxOff

\hypersetup{
    colorlinks=true,
    linkcolor=blue!40!black,
    urlcolor=green!40!black,
    citecolor=red!40!black,
    pdftitle={On the nature of the QCD chiral phase transition with imaginary chemical potential},
    pdfsubject={lqcd, columbia plot},
    pdfauthor={Alfredo D'Ambrosio, Michael Fromm, Reinhold Kaiser, Owe Philipsen},
}

\newcommand{\eq}{eq.~}

\newcommand{\fig}{figure~}
\newcommand{\figs}{figures~}
\newcommand{\refer}{ref.~}
\newcommand{\refers}{refs.~}
\newcommand{\tab}{table~}
\newcommand{\tabs}{tables~}
\newcommand{\sect}{section~}

\newcommand{\mr}{\multirow}
\newcommand{\mc}{\multicolumn}
\newcommand{\clqcd}{\texttt{CL\kern-.25em\textsuperscript{2}QCD}}
\newcommand{\bahamas}{\texttt{BaHaMAS}}
\newcommand{\amc}{am_\mathrm{c}}
\newcommand{\tc}{T_\mathrm{c}}
\newcommand{\betac}{\beta_\mathrm{c}}
\newcommand{\betapc}{\beta_\mathrm{pc}}
\newcommand{\betatric}{\beta_\mathrm{tric}}
\newcommand{\attric}{(aT)_\mathrm{tric}}
\newcommand{\atc}{(aT)_\mathrm{c}}
\newcommand{\nt}{N_\tau}
\newcommand{\nttric}{N_\tau^\mathrm{tric}}
\newcommand{\ns}{N_\sigma}
\newcommand{\nf}{N_\mathrm{f}}
\newcommand{\nfc}{N_\text{f}^\mathrm{c}}
\newcommand{\nftric}{N_\mathrm{f}^\mathrm{tric}}
\newcommand{\chisq}{\chi^2_\mathrm{d.o.f.}}

\newcommand{\botrule}{\bottomrule}

\definecolor{nfsixcolor}{HTML}{cc7700}
\definecolor{nffivecolor}{HTML}{800080}
\definecolor{nffourcolor}{HTML}{00a4b2}
\definecolor{nfsixcolorlight}{HTML}{ffbe66}
\definecolor{nffivecolorlight}{HTML}{ff1aff}
\definecolor{nffourcolorlight}{HTML}{4cf0ff}
\definecolor{ntfourcolor}{HTML}{0f2e8a}
\definecolor{ntsixcolor}{HTML}{0d7237}
\definecolor{nteightcolor}{HTML}{a01212}
\definecolor{ntfourcolorlight}{HTML}{7594f0}
\definecolor{ntsixcolorlight}{HTML}{47eb8a}
\definecolor{nteightcolorlight}{HTML}{f28d8d}

\title{
    On the nature of the QCD chiral phase transition with imaginary chemical potential
}

\author[1,2]{Alfredo D'Ambrosio\thanks{ambrosio@itp.uni-frankfurt.de}}
\author[1]{Michael Fromm\thanks{mfromm@itp.uni-frankfurt.de}}
\author[1,2]{Reinhold Kaiser\thanks{kaiser@itp.uni-frankfurt.de}}
\author[1,2]{Owe Philipsen\thanks{philipsen@itp.uni-frankfurt.de}}

\affil[1]{
    Institute for Theoretical Physics,
    Goethe-Universität Frankfurt,
    Max-von-Laue-Str. 1, 60438 Frankfurt am Main, Germany
}
\affil[2]{
    John von Neumann Institute for Computing (NIC),
    GSI Helmholtzzentrum für Schwerionenforschung,
    Planckstr. 1, 64291 Darmstadt, Germany
}
\date{}

\begin{document}
    \maketitle
    
	\begin{abstract}
        \input{abstract.tex}
    \end{abstract}

	\section{Introduction}\label{sec:introduction}
	\input{introduction.tex}

	\section{The Columbia plot at zero and imaginary chemical potential}\label{sec:theory}
	\input{theory.tex}

	\section{Lattice simulations and analysis}\label{sec:simulations}
	\input{simulations.tex}

	\section{Results}\label{sec:results}
	\input{results.tex}

	\section{Conclusions}\label{sec:conclusions}
	\input{conclusions.tex}

	\section*{Acknowledgements}
	\input{acknowledgments.tex}

	\appendix

	\section{Simulation statistics}\label{sec:statistics}
	\input{statistics.tex}

    \bibliographystyle{unsrturl}
    \bibliography{main}
\end{document}

%% file: abstract.tex
The order of the thermal chiral phase transition in lattice QCD is known to be strongly 
cutoff-dependent.
A previous study using  $\nf\in[2,6]$ mass-degenerate, unimproved staggered quark flavours on 
$\nt\in\{4,6,8\}$ lattices found that the bare mass regions displaying explicit first-order
transitions shrink to zero, with their critical boundary line terminating in a tricritical point
before the continuum limit is reached.
Here we perform an analogous study for fixed imaginary baryon chemical potential and find the
same behaviour:
first-order regions observed on coarse lattices disappear in tricritical points with diminishing
lattice spacing.
These observations are consistent with currently available results from improved staggered
discretisations, both at zero and non-zero imaginary chemical potential.
Unless additional first-order transitions are found on finer lattices or with chiral lattice
actions, this implies a second-order transition in the continuum chiral limit for all these
cases, at zero and imaginary chemical potential.
Implications for the $\nf=2+1$ QCD phase 
diagram at the physical point are discussed.

%% file: introduction.tex
The chiral limit of QCD, corresponding to massless quarks, is crucial for our understanding of
the strong interactions, because the physical $u,d$-quarks represent a small distortion of that situation.  
A natural question then is how the chiral crossover observed at physical quark masses~\cite{Aoki:2006we} emerges
from a necessarily non-analytic chiral phase transition in the limit $m_{u,d}\rightarrow 0$. Furthermore, the
nature of the chiral phase transition in the massless limit constrains the phase diagram of physical
QCD at finite temperature $T$ and baryon chemical 
potential~$\mu_B$~\cite{Rajagopal:1995bc,Halasz:1998qr,Hsu:1998eu,Stephanov:1998dy,Rajagopal:2000wf,Hatta:2002sj}.
Lattice QCD in the chiral limit cannot be simulated directly, because 
fermion matrix inversion becomes singular in that case. Therefore, extrapolations of simulation sequences with 
decreasing quark masses are necessary. 
Despite these difficulties,
decisive progress has been achieved over the last few years, as we now summarise.

The quark mass dependence of
the pseudo-critical cross\-over temperature appears to be only weakly sensitive to the precise set of critical exponents,
with which the chiral limit is approached, or even to the order of the transition. 
This allows for stable extrapolations $m_{u,d}\rightarrow 0$ at the physical strange
quark mass, leading to $\tc(m_{u,d}=0,m_s^\mathrm{phys})=132^{+3}_{-6}\unit{MeV}$ when using highly improved staggered (HISQ)
fermions~\cite{Ding:2019prx}, and    
$\tc(m_{u,d}=0,m_s^\mathrm{phys})=134^{+6}_{-4}\unit{MeV}$ when using twisted mass Wilson fermions~\cite{Kotov:2021rah}. 

\begin{figure}[t]
    \centering
    \begin{subfigure}[t]{0.49\linewidth}
        \centering
        \includegraphics[width=\linewidth]{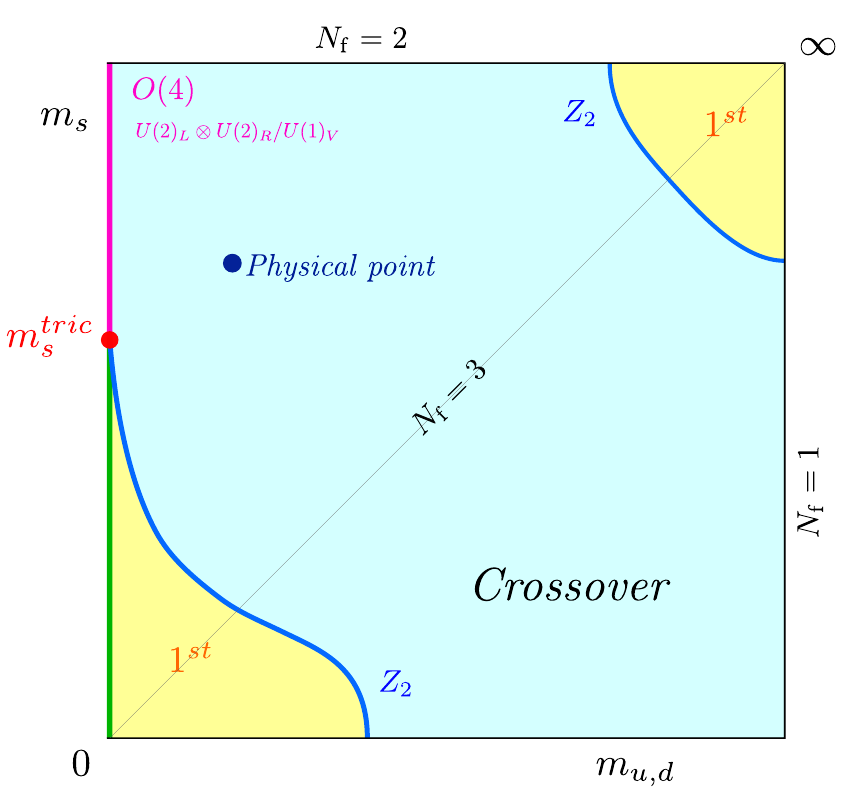}
        \caption{
            Prediction by 3d $\phi^4$-theories, augmented by a 't Hooft 
            determinant~\cite{Pisarski:1983ms,Pelissetto:2013hqa}.
            Figure taken from \refer\cite{Cuteri:2021ikv}.
        }\label{fig:columbia_1st}
    \end{subfigure}
    \hfill
    \begin{subfigure}[t]{0.49\linewidth}
        \centering
        \includegraphics[width=\linewidth]{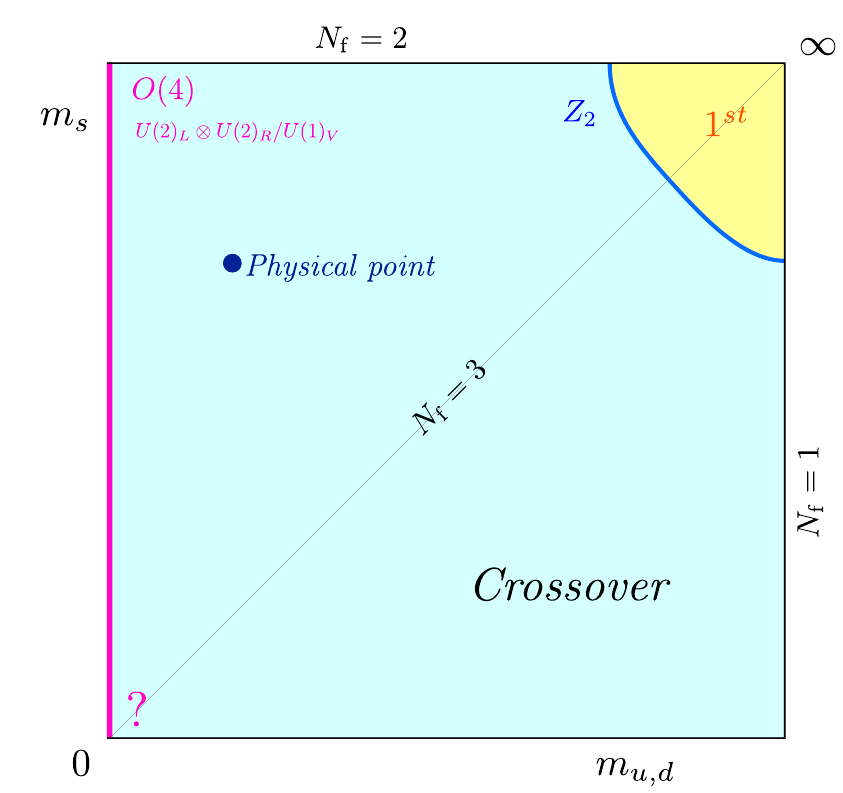}
        \caption{
            Prediction by lattice QCD with standard staggered fermions~\cite{Cuteri:2021ikv}, 
            consistent with all presently available lattice results.
            Figure taken from \refer\cite{Cuteri:2021ikv}.
        }\label{fig:columbia_2nd}
    \end{subfigure}
    \caption{
        Nature of the $\nf=2+1$ QCD thermal transition as a function of quark masses.
        Figures taken from~\cite{Cuteri:2021ikv}.
    }\label{fig:columbia}
\end{figure}

By contrast, it is much harder to obtain robust lattice results on the order of the chiral transition because
of the necessary finite size scaling analysis, in addition to the continuum and chiral extrapolations.
For $\nf=2+1$ QCD, the nature of the thermal transition as a function of quark masses
is schematically represented in the so-called Columbia plot, \figs\ref{fig:columbia_1st} 
and~\ref{fig:columbia_2nd}. 
For more than 30 years, the version displayed in \fig\ref{fig:columbia_1st} was expected to be realised by QCD, 
based on renormalisation group
(RG) flows in linear sigma models~\cite{Pisarski:1983ms,Gausterer:1988fv,Butti:2003nu,Pelissetto:2013hqa} and
early numerical results obtained on coarse lattices~\cite{Brown:1990ev,Iwasaki:1996zt}. 
However, the chiral critical $Z_2$-line~\cite{Karsch:2001nf,deForcrand:2003vyj} delimiting the first-order region was found to be 
strongly cutoff-dependent, with enormous apparent differences between different lattice actions, 
see~\cite{Philipsen:2021qji,Lahiri:2021lrk,Guenther:2022wcr} for reviews and 
references.

A resolution was suggested by mapping out the
chiral critical boundary between first-order and crossover regions 
in an enlarged multi-flavour parameter space using 
unimproved\footnote{The primary interest of our studies is to understand cutoff effects on the order
of the phase transition. It is therefore beneficial to work with a discretisation that displays them clearly.} 
staggered fermions~\cite{Cuteri:2021ikv}. All theories with $\nf\in[2,6]$ mass-degenerate flavours exhibit a 
first-order chiral transition region on coarse $\nt=4$ lattices, which shrinks with increasing $\nt$ (viz.~decreasing lattice spacing)
to disappear in a tricritical point, before the
continuum limit is reached. This implies that the continuum chiral limit 
corresponds to a second-order transition for all these cases. Complementary to this, a re-analysis of previously published data for
$\nf=3$ $\order{a}$-improved Wilson fermions~\cite{Kuramashi:2020meg} was also found to be consistent with tricritical scaling 
and thus a second-order transition in the continuum limit~\cite{Cuteri:2021ikv}. 
Unless an additional first-order region emerges in future simulations on finer lattices or with chiral fermions, 
these results suggest a continuum 
Columbia plot as in \fig\ref{fig:columbia_2nd}. Note that currently nothing is known about the universality class
in the lower left $\nf=3$ corner, whose classical chiral symmetry is $U(3)\times U(3)$. 

Since a $\nf=3$ second-order transition differs qualitatively from earlier predictions, independent checks are crucial.
Investigations of the $\nf=3$ theory with two further 
discretisations are also consistent with these results: no sign of a 
first-order phase transition is detected with HISQ fermions down to $m_\pi\approx 45\unit{MeV}$~\cite{Dini:2021hug},
nor with domain wall fermions at a mass value of the physical 
$u,d$-quarks~\cite{Zhang:2022kzb,Zhang:2025vns,Zhang:2024ldl}. 
In \refer\cite{Fejos:2022mso} the RG flow of a three-dimensional $\phi^6$-theory was studied with functional methods,
finding an infrared fixed point as a candidate for a second-order transition, depending on the fate of the $U(1)_A$ anomaly
at the transition temperature. Conditions on the anomaly permitting a second-order transition are also formulated in~\cite{Pisarski:2024esv,Giacosa:2024orp}.
A similar RG analysis predicts second-order transitions for $\nf\geq 5$ independent of the anomaly~\cite{Fejos:2024bgl}.
Numerical conformal bootstrap methods 
provide evidence for the possibility of a second-order transition in generic
$U(3)\times U(3)$ models~\cite{Kousvos:2022ewl}. 
A second-order chiral transition is also reported in $\nf=2+1$ QCD for all strange quark mass values using truncated Dyson-Schwinger 
equations (DSE) of continuum QCD, where the massless limit can be studied explicitly~\cite{Bernhardt:2023hpr}. 

In this paper we continue to study the predictions of standard staggered fermions, 
and investigate how the order of the chiral phase
transition depends on baryon chemical potential $\mu_B$.
As a first step, we repeat the analysis of \refer\cite{Cuteri:2021ikv} for a fixed imaginary chemical potential, for which 
there is no sign problem. The emerging qualitative picture is completely analogous to the situation at $\mu_B=0$:
while a first-order phase transition region is explicitly seen on coarse $\nt=4$-lattices~\cite{deForcrand:2003vyj}, it 
shrinks with diminishing lattice spacing to disappear in a tricritical line.\footnote{Preliminary evidence has been given  
in \refer\cite{DAmbrosio:2022kig}.} Again this is consistent with the picture from DSE studies~\cite{Bernhardt:2023hpr,Bernhardt:2025fvk},
where the nature of the chiral transition was found to be independent of imaginary or small real chemical potentials.
Moreover, the DSEs in their current truncation explicitly demonstrate analyticity
of the chiral critical surface around $\mu_B=0$, thus further motivating 
lattice studies at imaginary chemical potentials.

%% file: theory.tex
\begin{figure}[t]
    \centering
    \includegraphics[width=0.5\linewidth]{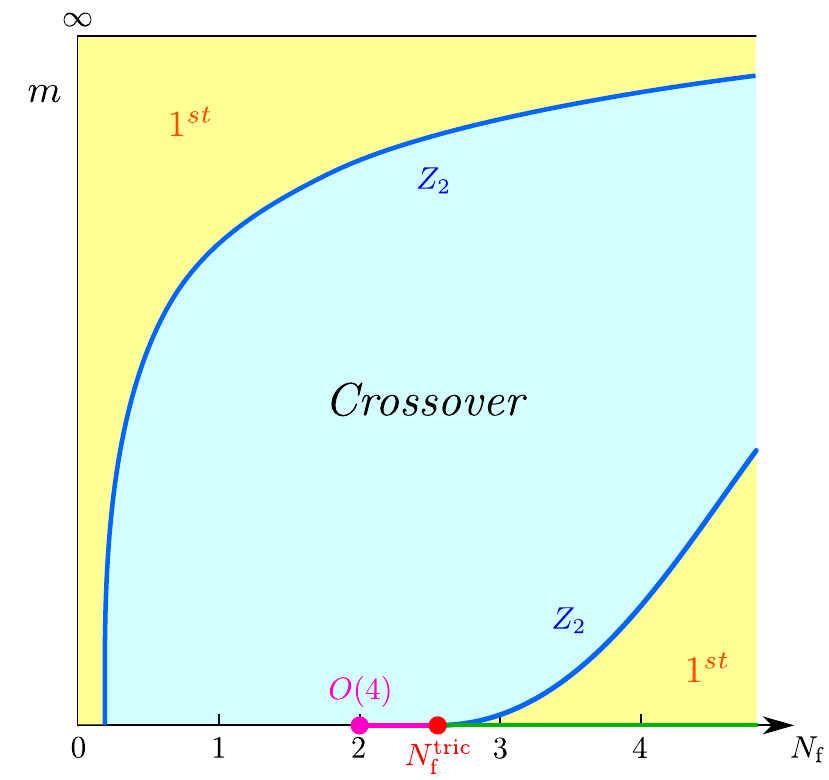} 
    \caption{
        Columbia plot for mass-degenerate quarks, scenario analogous to \fig\ref{fig:columbia_1st}.
        Figure taken from \refer\cite{Cuteri:2021ikv}.
    }\label{fig:nf_columbia}
\end{figure}

To render this paper self-contained, we briefly summarise the theoretical concepts for our numerical
strategy, which have been first proposed and discussed in more detail in \refers\cite{Cuteri:2017gci,Cuteri:2021ikv}.
Instead of $\nf=2+1$ QCD we consider a parameter space with a continuously variable 
number $\nf$ of mass-degenerate quarks with bare mass $m$.
Just as the strange quark mass $m_s\in [m_{u,d}=m,\infty)$ interpolates between theories with $\nf=3$ or $\nf=2$ flavours of
mass $m$, respectively, so does a continuous variation of the power $\nf\in [2,3]$, to which the quark
determinant is raised in the QCD path integral,
\begin{equation}
    Z(\nf,g,m)=\int {\cal D}A_\mu \; (\det M[A_\mu,m])^{\nf}\; e^{-S_\text{YM}[A_\mu]}\;.
\end{equation}  
Such a mapping $\{m_{u,d},m_s\}\rightarrow \{m,\nf\}$ is not unique, but that is not important for our purposes. In practice, we work 
with rooted staggered fermions where the desired roots can be straightforwardly adapted.
The Columbia plot version from \fig\ref{fig:columbia_1st} then changes
to \fig\ref{fig:nf_columbia}, which features several advantages: 
\begin{itemize}
    \item Any first-order transition region observed at sufficiently large $\nf$ must disappear in a tricritical
    point $\nftric$, because there is no chiral phase transition for integer $\nf<2$. 
    \item The $Z_2$ boundary line delimiting the first-order transition region then represents a so-called wing
    line, which enters the tricritical point as a function of the symmetry breaking scaling field with known
    mean-field exponents~\cite{lawrie},
    \begin{equation}\label{eq:scale}
        \nfc(m)=\nftric + A \cdot m^{2/5} + B \cdot m^{4/5}\;.
    \end{equation}
\end{itemize}
Starting from a known $Z_2$ boundary point, one can map out the chiral critical line
towards smaller masses and extrapolate it in a controlled way to determine the location of $\nftric$.
One would then conclude that all integer $\nf>\nftric$ feature first-order and all integer $\nf<\nftric$ feature 
second-order chiral transitions.

On the lattice, the parameter space gets enlarged by the lattice spacing. Hence, the $Z_2$-boundary separating
first-order transitions from crossover represents a surface, which in the lattice chiral limit ($am=0$) terminates in a tricritical line
$\nftric(a)$. Using $T=(a \nt)^{-1}$,  this can be expressed as $\nftric(\nt)$. 
In \refers\cite{Cuteri:2017gci,Cuteri:2021ikv} this
line was bounded in the space of $\nf\in[2,6]$ and $\nt\in\{4,6,8\}$. 
The results imply a possible tricritical point in the continuum to be at 
$\nftric(\nt\rightarrow\infty) > 6$.
Inverting this relationship to 
$\nttric(\nf)$, one can restrict to the physical case of integer $\nf$ again.
The statement then is that for $\nf\in[2,6]$ there exists a finite  $\nttric(\nf) < \infty$.
This means that the first-order chiral transition is not analytically 
connected to the continuum limit, which then represents a second-order transition.
Returning to the $\nf=2+1$ situation,
one thus arrives at the continuum scenario displayed in \fig\ref{fig:columbia_2nd}.

\begin{figure}[h]
    \centering
    \includegraphics[width=0.55\linewidth]{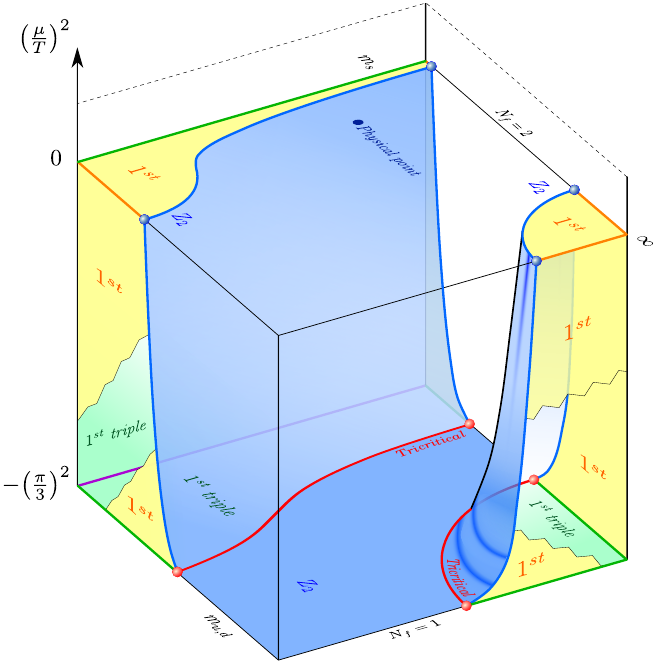}
    \caption{
        Columbia plot with chemical potential as observed on coarse lattices. 
        The bottom plane corresponds to the first Roberge-Weiss 
        center-transition~\cite{Roberge:1986mm}.
        Taken from~\cite{Sciarra:2016exn,Philipsen:2019ouy}.
    }\label{fig:3dcolumbia}
\end{figure}

In the presence of a quark chemical potential, $\mu=\mu_B/3$, the Columbia plot (in the continuum or at fixed lattice spacing) 
gets extended into a third dimension. In this case the $Z_2$-critical boundary lines 
sweep out critical surfaces, as sketched schematically in \fig\ref{fig:3dcolumbia}, which depicts the situation 
found on coarse lattices. Labelling the third axis
by $(\mu/T)^2$, both real and imaginary chemical potential can be discussed.  
Imaginary chemical potential, $\mu=i\mu_i$  with $\mu_i\in \mathbb{R}$, is unphysical, but it does not induce a sign
problem and the phase structure can be simulated without difficulty or further approximations.
For arbitrary fermion masses the partition function
is periodic in $\mu_i$ according to the global Roberge-Weiss (or center) 
symmetry~\cite{Roberge:1986mm},
\begin{equation}
    Z\left(T,i\frac{\mu_i}{T}\right)=Z\left(T,i\frac{\mu_i}{T}+i\frac{2\pi n}{N_c}\right)\;.
\end{equation}
The boundaries between different center sectors are located at
\begin{equation}
(\mu/T)_c=\pm i(2n+1)\pi /3\;,
\end{equation}
with $n=0,1,2,\ldots$, with the bottom plane of \fig\ref{fig:3dcolumbia}
representing the first of these. All adjacent sectors are related to the first one by symmetry.

For unimproved Wilson~\cite{Philipsen:2016hkv,Cuteri:2015qkq} and 
staggered~\cite{deForcrand:2010he,PhysRevD.83.054505,Philipsen:2019ouy} discretisations on $\nt\in\{4,6\}$, 
the 3D~Columbia plot looks as in  
\fig\ref{fig:3dcolumbia}, with the region of chiral phase transitions getting wider in the imaginary
$\mu$ direction. Expanding the $Z_2$-critical light mass for small chemical potential for any fixed $m_s$,
\begin{equation}
    \label{eq:m_cont}
    \frac{m^\mathrm{c}_l(\mu)}{m^\mathrm{c}_l(0)}=1+c_1 \left(\frac{\mu}{T}\right)^2+\order{\left(\frac{\mu}{T}\right)^4}\;,
\end{equation}
one may conclude that $c_1<0$ and by analytic continuation the first-order region shrinks in the real-$\mu$ direction.
As an independent check on these calculations at imaginary chemical potential, the curvature of the chiral
critical surface can be computed directly at $\mu=0$. 
For staggered fermions one finds $c_1<0$ both at $m_s=m_l$ and at $m_s=m_s^\mathrm{phys}$,
and the next coefficient is negative as well~\cite{deForcrand:2006pv,deForcrand:2007rq,deForcrand:2008vr}.
The only calculation finding a positive curvature so far is based on $\order{a}$-improved Wilson fermions~\cite{Jin:2015taa}.

Investigations in the Roberge-Weiss plane on finer lattices reveal the same trend as seen 
at $\mu=0$, namely the chiral tricritical line moving towards
smaller quark masses, both for unimproved staggered~\cite{Philipsen:2019ouy} and Wilson~\cite{Cuteri:2015qkq} quarks.
On the other hand, for stout-smeared staggered~\cite{Bonati:2018fvg} and HISQ~\cite{Cuteri:2022vwk} actions, 
a first-order region in the Roberge-Weiss plane cannot be detected even on $\nt=4$, when starting from the physical point and
reducing the pion masses down to $m_\pi\approx 50\;\unit{MeV}$.

In the following sections, we study how the $Z_2$ boundary of the first-order transition region at imaginary chemical potential
changes as a function of $\nf$ and the lattice spacing using unimproved staggered fermions, 
i.e., we repeat the analysis of \refer\cite{Cuteri:2021ikv} for a 
fixed chemical potential of $\mu_i=0.81 \pi T/3$. 
This value of $\mu_i$ is chosen to maximise the effect of chemical potential by a large value,
while remaining sufficiently distant
from the Roberge-Weiss plane, thereby avoiding unwanted critical behaviour associated with 
the Roberge-Weiss transition.

%% file: simulations.tex
For all Monte Carlo simulations we used the standard Wilson plaquette action for the gauge sector and the 
unimproved staggered action for the fermion sector.
Gauge configurations were generated with the code \clqcd\, version~\texttt{v1.1}~\cite{sciarra_cl2qcd_2021}, 
which employs the RHMC algorithm for degenerate quark flavours.
The multiple pseudo-fermion technique was used to reduce the noise from stochastic estimates of the fermion determinant,
with the optimal number of pseudo-fermions calculated by \clqcd\ at the beginning of a run.
The simulations were run on the Virgo cluster at GSI (Darmstadt) on AMD GPUs of type MI100,
and monitored efficiently using the bash-software \bahamas~\cite{sciarra_bahamas_2021}.
In total almost $150$~million trajectories have been generated, spread over about $600$ different parameter sets.
For each  parameter set, four differently seeded Markov chains
were run independently to increase statistics. Our analysis follows the same steps as in~ref.~\cite{Cuteri:2021ikv}.

In order to determine the location and order of the chiral transitions, we use the chiral condensate as (quasi-)order parameter 
$O = \bar\psi\psi$, which we measure for each trajectory 
using 16 stochastic estimators. We then analyse its generalised moments,
\begin{equation}
	\label{equ:std-moments}
	B_n(\beta, am , \ns) = \frac{\ev{\left( O - \ev{O}\right)^n}}{\ev{\left(O - \ev{O}\right)^2}^{n/2}}\;.
\end{equation}
When calculating the higher moments of the $\bar\psi\psi$, the bias subtraction method is applied,
which prevents estimators of the $\bar\psi\psi$ being multiplied with themselves.

\begin{table*}[t]
	\caption{
		Fit results for the critical quark mass and the associated gauge coupling.
	}\label{tab:amc-results}
	\input{amc-betac-results.tex}
\end{table*}

In the four dimensional parameter space spanned by the lattice gauge coupling $\beta$, the quark mass $am$, the number of quark flavours $\nf$ and the temporal lattice extent $\nt$, the chiral first-order region is separated from the crossover region by a $Z_2$-second-order surface.
In practice we first scan in the lattice gauge coupling to determine the (pseudo-)critical hyper-surface (the ``phase boundary'')
defined by vanishing skewness, $B_3(\betapc, am, \ns)=0$,
and then employ the kurtosis  evaluated on the phase boundary, $B_4(\betapc,am,\ns)$ 
in order to identify the chiral critical $Z_2$-surface in this subspace.
In order to gain precision for $\betapc$, we interpolate both standardized moments between
the simulated $\beta$-values according to the multiple histogram method~\cite{Ferrenberg:1988yz}.
The $Z_2$-critical mass is identified by the kurtosis assuming its associated critical 
value of $B_4^{Z_2}=1.6044(10)$~\cite{Blote:1995zik}. In practice, we fit the kurtosis in the neighbourhood
of the critical mass value  to a finite size scaling formula,
\begin{equation}
	\label{equ:kurtosis-fss}
		B_4(\betapc,am,\ns)\approx
		\left(1.6044 + c(am-\amc)\ns^{1/\nu}\right)\left(1+b\ns^{y_t-y_h}\right)\;,
\end{equation}
which gives the critical mass $\amc$ as a fit parameter.
The last factor on the right hand side of \eq\eqref{equ:kurtosis-fss} contains a finite volume correction term, 
where $y_t=1/\nu=1.5870(10)$ and $y_h=2.4818(3)$~\cite{Pelissetto:2000ek} are the associated 3D Ising exponents.
This term becomes statistically insignificant for sufficiently large volumes.

The values we obtained for the critical mass and coupling, $\amc$ and $\betac$, respectively, 
are reported in \tab\ref{tab:amc-results}, which also includes fitting details.
The value of $\betac$ is obtained 
from a linear fit of the pseudo-critical $\beta$-values at the simulated quark masses.
Evaluating the obtained fit function also at $\amc\pm\sigma_{\amc}$ leads to the attached error of $\betac$.
A detailed overview of the statistics associated to the results from \tab\ref{tab:amc-results} are
given in appendix~\ref{sec:statistics}.

%% file: amc-betac-results.tex
\begin{tabular*}{\textwidth}{@{\extracolsep\fill}lllllllll}
    \toprule
    $\nt$           & $\nf$ & $am_\text{min}$   & $am_\text{max}$	& $\amc$	    & fit type	    & d.o.f.	& $\chisq$	& $\betac$ at $\amc$	\\
    \midrule
    \mr{11}{*}{4}   & 1.8   & 0.0014            & 0.0050            & 0.0023(8)     & linear+corr   & 13        & 0.34      & 5.3162(13)            \\
                    & 1.9	& 0.0020		    & 0.0060			& 0.0032(5)		& linear    	& 6			& 0.13 	    & 5.3007(8)		     	\\
                    & 2.0	& 0.0040		    & 0.0120			& 0.0054(6)		& linear     	& 7			& 0.26 		& 5.2879(11) 			\\
                    & 2.1	& 0.0060		    & 0.0130			& 0.0079(6)		& linear    	& 7			& 0.20   	& 5.2757(11) 			\\
                    & 2.2	& 0.0080		    & 0.0140			& 0.0098(6)		& linear		& 7			& 0.43		& 5.2628(11)			\\
        		    & 2.3	& 0.0065		    & 0.0230			& 0.0126(5)		& linear		& 9			& 0.59		& 5.2517(9)				\\
                    & 3.6	& 0.0370		    & 0.0520			& 0.0494(11)	& linear+corr	& 9 		& 0.24		& 5.1315(20)			\\
                    & 4.0	& 0.0500		    & 0.0650			& 0.0596(10)	& linear+corr	& 9 		& 0.08		& 5.0993(18)			\\
                    & 4.5	& 0.0650		    & 0.0750			& 0.0731(11)	& linear+corr	& 6			& 0.19		& 5.0635(20)			\\
                    & 5.0	& 0.0750		    & 0.1050			& 0.0872(11)	& linear+corr	& 9			& 0.23		& 5.0324(20)			\\
                    & 6.0	& 0.1000		    & 0.1300			& 0.1119(12)	& linear+corr	& 6			& 0.04		& 4.9732(23)			\\
    \midrule
    \mr{7}{*}{6}    & 3.0	& 0.0010		    & 0.0030			& 0.0021(3)		& linear+corr	& 5			& 0.79		& 5.2292(13)			\\
                    & 3.3	& 0.0030		    & 0.0075			& 0.00433(20)	& linear+corr	& 9			& 0.25		& 5.1877(9) 			\\
                    & 3.6	& 0.0050		    & 0.0100			& 0.0069(3)		& linear+corr	& 6			& 0.15		& 5.1490(13)			\\
         		    & 4.0	& 0.0075		    & 0.0125			& 0.0107(3)		& linear+corr	& 6 		& 0.31		& 5.1010(14)			\\
                    & 4.5	& 0.0100		    & 0.0180			& 0.0151(4)		& linear+corr	& 6			& 0.95		& 5.0425(16)			\\
                    & 5.0	& 0.0175		    & 0.0250			& 0.02082(25)	& linear+corr	& 6			& 0.59		& 4.9913(10)			\\
                    & 6.0	& 0.0300		    & 0.0380			& 0.0320(4)		& linear+corr	& 6			& 0.60		& 4.8960(14)			\\
    \midrule
    \mr{5}{*}{8}    & 4.0	& 0.0010		    & 0.0030			& 0.00164(26)   & linear+corr	& 8			& 0.66		& 5.1122(18)			\\
                    & 4.5	& 0.0020		    & 0.0040     		& 0.00309(18)	& linear+corr	& 5			& 0.11		& 5.0367(14)  			\\
        		    & 5.0	& 0.0040		    & 0.0075			& 0.00499(17)	& linear+corr	& 8 		& 0.98		& 4.9676(12)			\\
                    & 5.5	& 0.0050		    & 0.0075			& 0.00663(17)	& linear+corr   & 5			& 0.23		& 4.8990(12)			\\
                    & 6.0	& 0.0060		    & 0.0100			& 0.00879(19)	& linear+corr	& 5			& 0.01		& 4.8359(13)			\\
    \botrule
\end{tabular*}

%% file: results.tex
The simulations and analysis described in the previous section provides us with the location of the
chiral critical surface in the bare parameter space $\{\beta,am,\nf,\nt\}$ of standard staggered fermions,
with fixed $\mu_i=0.81\pi T/3$. We will now project this critical surface on planes defined by different variable pairings
and discuss the implications.

\begin{figure}[t]
    \begin{subfigure}[t]{0.49\linewidth}
        \centering
        \includegraphics[scale=0.99]{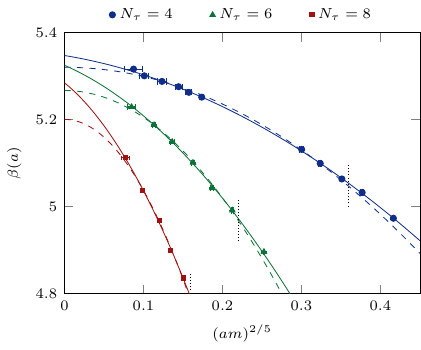}
        \caption{
            $\betac$ as function of the rescaled bare quark mass $(am)^{2/5}$ for several
            staggered flavours $\nf$ at varying temporal extent $\nt$.
            The tricritical scaling fit (\eq\eqref{equ:beta_c} and \tab\ref{tab:beta_c-fit-results})
            includes LO+NLO (solid) and NLO (dashed) terms, respectively.
            The dotted bars mark the upper limit of critical masses included in the respective
            fits.
        }\label{fig:beta_c} 
    \end{subfigure}
    \hfill
    \begin{subfigure}[t]{0.49\linewidth}
        \centering
        \includegraphics[scale=0.99]{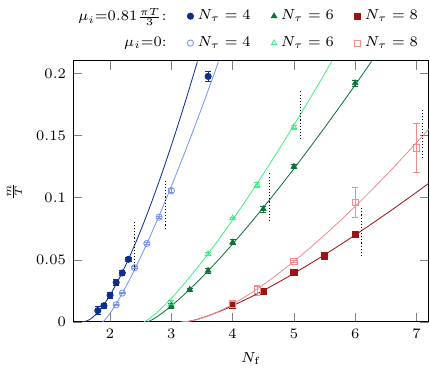}
        \caption{
            The critical curves in the plane spanned by the quark mass and $\nf$.
            Fits correspond to \eq\eqref{equ:nf_c} with the fit results in
            \tab\ref{tab:am-nf-fit-results}.
            Comparison between the data from $\mu_i=0$~\cite{Cuteri:2021ikv} and
            $\mu_i=0.81\frac{\pi T}{3}$.
            The dotted bars mark the upper limit of critical masses included in the respective
            fits.
        }\label{fig:m_t-nf} 
    \end{subfigure}
    \caption{
        Projections of the critical surface on the $(\beta,am)$-plane (left) and on the 
        $(\nf,m/T)$-plane (right).
    }
\end{figure}

\begin{table*}[t]
    \caption{
        Fits of \eq\eqref{equ:beta_c} to the data in \fig\ref{fig:beta_c}.
    }\label{tab:beta_c-fit-results}
    \input{beta-am-fit-coeffs-table}
\end{table*}

\begin{table*}[t]
    \caption{
        Fits of \eq\eqref{equ:nf_c} to the data for $\mu_i=0.27\pi T$ in \fig\ref{fig:m_t-nf}.
    }\label{tab:am-nf-fit-results}
    \input{am-nf-fit-coeffs-table}
\end{table*}

Beginning in the $(\beta,am)$-plane, \fig\ref{fig:beta_c} shows the $Z_2$-critical boundary line separating the parameter region
with crossover to the right of it from the first-order region to the left of it, for lattices with different $\nt$. The different
data points are implicitly parametrised by different values of $\nf$, which is increasing from left to right. 
No difference between integer and non-integer $\nf$-values is discernible.
The lines show fits to leading and next-to-leading order
in the tricritical scaling field.
For all three $\nt$-values, the critical curves are well described over a wide mass range by a next-to-leading-order fit in the scaling variable,
cf.~\tab\ref{tab:beta_c-fit-results},
\begin{equation}
    \label{equ:beta_c}
        \betac(am,\nf(\nt),\nt) =
        \betatric(\nt) + \mathcal{C}_1(\nt)(am)^{2/5} + \mathcal{C}_2(\nt)(am)^{4/5}
        + \order{(am)^{6/5}}\; .
\end{equation}
In \fig\ref{fig:beta_c}, the solid lines correspond to fits including leading-order and
next-to-leading-order terms (LO+NLO).
Dashed lines correspond to fits that do not include the leading-order term and only include
the next-to-leading-order term (NLO).
As explained in \sect\ref{sec:theory}, a projection of the chiral critical surface onto a plane containing $\nf$ must
necessarily terminate in a tricritical point. For this reason the fit ansatz used here is justified, even though the leading term
is just beginning to be constrained.\footnote{It is easy to check that a description by a polynomial in terms of $am$ instead of the
scaling field is indeed significantly worse, quantitatively.}
Compared to the next-to-leading-order term, the leading-order term is small.
As a result, only computationally expensive simulations at even lower masses would constrain 
the leading-order term.

Next, the $(am,\nf)$-plane is shown in \fig\ref{fig:m_t-nf},
which corresponds to the schematic picture from \fig\ref{fig:nf_columbia}. 
The lines correspond to fits according to
the scaling relation,
\begin{equation}
    \label{equ:nf_c}
        \nfc(am,\nt) =
        \nftric(\nt) + \mathcal{D}_1(\nt)(am)^{2/5} + \mathcal{D}_2(\nt)(am)^{4/5}
        + \order{(am)^{6/5}}\;.
\end{equation}
For comparison, we have also included 
the previous $\mu_i=0$ data from \refer\cite{Cuteri:2021ikv} in this figure. We observe that in both cases the tricritical point, in which
the critical line terminates, is strongly cutoff-dependent and moves towards larger $\nf$-values as the lattice is made finer,
cf.~\tab\ref{tab:am-nf-fit-results}. Given this qualitative similarity, there is an interesting quantitative detail in the comparison.
On the coarsest $\nt=4$ lattices, the first-order region for finite $\mu_i$ extends to larger bare quark masses than in the
case $\mu_i=0$, in agreement with earlier studies using staggered fermions on  
$\nt=4$~\cite{deForcrand:2006pv,deForcrand:2007rq,deForcrand:2008vr}. However, on the $\nt\in\{6,8\}$ lattices this situation
changes and the $\mu_i=0$ data show the larger first-order region, which is consistent with the 
Wilson study on $\nt=6$ quoted earlier~\cite{Jin:2015taa}.
This is clear evidence for a mixing of cutoff and finite density effects  in the parameter region where we are operating, which is not surprising: we are not yet close to the continuum, nor is our chosen $\mu_i/T$-value small.

\begin{figure}[t]
    \begin{subfigure}[t]{0.49\linewidth}
        \centering
        \includegraphics[scale=0.99]{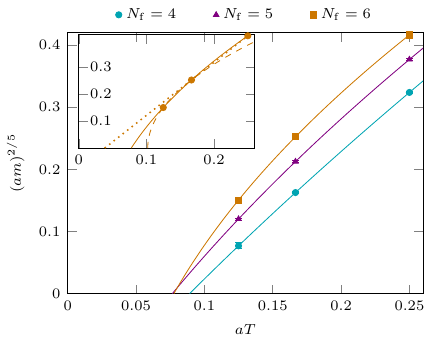}
        \caption{
            Tricritical scaling interpolations (\eq\eqref{eq:T_m}). 
            The inset shows LO (dotted) and NLO (dashed) interpolations for $\nf=6$ for 
            an error estimate of $\attric$.
            The NLO interpolation does not include the leading-order term.
        }\label{fig:am-at-tric} 
    \end{subfigure}
    \hfill
    \begin{subfigure}[t]{0.49\linewidth}
        \centering
        \includegraphics[scale=0.99]{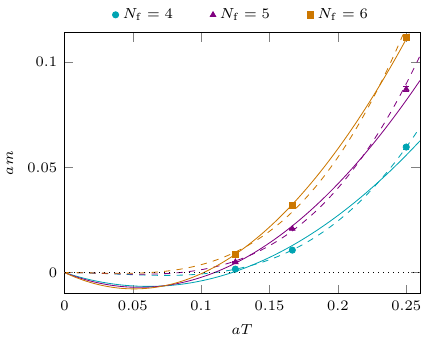}
        \caption{
            Polynomial fits.
            Solid lines correspond to LO+NLO fits and dashed lines to NLO+NNLO fits.
            Fits correspond to \eq\eqref{eq:m_T_1st} with the fit results in 
            \tab\ref{tab:am-at-fit-results-polynomial}.
        }\label{fig:am-at-polynomial}
    \end{subfigure}
    \caption{Critical line for different $\nf$ in the $(am,aT)$-plane.}
\end{figure}

\begin{table*}[h]
    \caption{
        Fit coefficients of the fits from \eq\eqref{eq:m_T_1st} to the data in 
        \fig\ref{fig:am-at-polynomial}.
    }\label{tab:am-at-fit-results-polynomial}
    \input{am-at-fit-coeffs-polynomial-table}
\end{table*}

Finally, we project the chiral critical surface on the $(am,aT=\nt^{-1})$-plane in
\fig\ref{fig:am-at-tric}.
Once again, the observed behaviour is 
completely analogous to the case at zero density.
The second-order boundary line between the first-order and crossover regions is well described by next-to-leading-order
tricritical scaling of temperature in terms of the scaling field,
\begin{equation}\label{eq:T_m}
        \atc(am,\nf) = 
        \attric(\nf) + E_1(\nf) (am)^{2/5} + E_2(\nf)(am)^{4/5} 
        + \order{(am)^{6/5}}\;.
\end{equation}
With only three $\nt$-values available so far, a full next-to-leading-order fit involving three fit
parameters is not feasible.
The lines in \fig\ref{fig:am-at-tric} thus represent exact interpolations using LO and NLO terms going 
exactly through the central values.
Note that for fixed $\nf$ there is no guarantee for a tricritical point to exist in this plane:
if the theory with a given $\nf$ features a first-order
transition in the continuum, the $Z_2$-boundary line must enter the continuum limit $(am,aT)=(0,0)$ without tricritical scaling, i.e.~as
an ordinary polynomial
\begin{equation}\label{eq:m_T_1st}
        \amc(\nt,\nf)= 
        \tilde{F}_1(\nf)\; aT+\tilde{F}_2(\nf)\;(aT)^2 + \tilde{F}_3(\nf)\;(aT)^3 
        + \order{(aT)^4}\;.
\end{equation}
As \fig\ref{fig:am-at-polynomial} and \tab\ref{tab:am-at-fit-results-polynomial} show, this functional behaviour is incompatible with the data. Instead, the data are
fully compatible with the tricritical scaling behaviour observed in the other planes of bare parameter pairings, and hence with 
the existence of a tricritical line $\nftric(\nt)$, viz.~$\nttric(\nf)$ in the lattice chiral limit, $am=0$.

Our findings imply that, just as for zero density, there is a maximal $\nt$ beyond which the first-order transition observed in our
simulations is lost, i.e., it is not connected to the continuum limit and thus must be considered a lattice artefact.
Since the tricritical point marks a change from first- to second-order behaviour in the lattice chiral limit, 
the continuum chiral limit corresponds to a second-order transition. The Columbia plot for $\mu_i=0.81 \pi T/3$ then looks
qualitatively the same as in \fig\ref{fig:columbia_2nd}.

%% file: beta-am-fit-coeffs-table.tex
\begin{tabular*}{\textwidth}{l@{\extracolsep{\fill}}lllllll}
	\toprule
	 $\nt$                 & $\mathcal{C}_2$       & $\mathcal{C}_1$       & $\betatric$           & range in $am$         & fit form              & d.o.f.                & $\chisq$              \\
	\midrule
	\multirow{2}{*}{4}	   & $-1.43(15)$           & $-0.30(7)$            & $5.347(7)$            & $[0,0.08]$            & LO+NLO                & $6$                   & $0.494$               \\
						   & $-2.11(5)$            & ---                   & $5.3204(24)$          & $[0,0.08]$            & NLO                   & $7$                   & $3.02$				   \\
	\hdashline						   
	\multirow{2}{*}{6}     & $-3.7(1.0)$           & $-0.8(3)$             & $5.325(24)$           & $[0,0.03]$            & LO+NLO                & $3$                   & $0.623$               \\
						   & $-6.17(14)$           & ---                   & $5.267(4)$            & $[0,0.03]$            & NLO                   & $4$                   & $1.65$                \\
	\hdashline
	\multirow{2}{*}{8}     & $-10(6)$              & $-1.5(1.5)$           & $5.28(9)$             & $[0,0.01]$            & LO+NLO                & $2$                   & $0.834$               \\
						   & $-16.3(6)$            & ---                   & $5.200(10)$           & $[0,0.01]$            & NLO                   & $3$                   & $0.860$               \\
	\botrule
\end{tabular*}

%% file: am-nf-fit-coeffs-table.tex
\begin{tabular*}{\textwidth}{l@{\extracolsep{\fill}}lllllll}
    \toprule
    $\nt$                 & $\mathcal{D}_2(\nt)$  & $\mathcal{D}_1(\nt)$  & $\nftric$             & range in $am$         & fit form              & d.o.f.                & $\chisq$              \\
    \midrule
    $4$                   & $13(8)$               & $2.1(2.2)$            & $1.54(15)$            & $[0,0.015]$           & LO+NLO                & $3$                   & $0.316$               \\
    $6$                   & $54(10)$              & $0.0(2.6)$            & $2.61(17)$            & $[0,0.018]$           & LO+NLO                & $2$                   & $0.345$               \\
    $8$                   & $11(4)\cdot 10^{1}$   & $2(9)$                & $3.2(5)$              & $[0,0.01]$            & LO+NLO                & $2$                   & $0.589$               \\
    \botrule
\end{tabular*}

%% file: am-at-fit-coeffs-polynomial-table.tex
\begin{tabular*}{\textwidth}{l@{\extracolsep{\fill}}llllll}
    \toprule
     $\nf$                 & $\tilde{F}_3(\nf)$    & $\tilde{F}_2(\nf)$    & $\tilde{F}_1(\nf)$    & fit form   & d.o.f.                & $\chisq$              \\
    \midrule
    \mr{2}{*}{$4.0$}       & ---                   & $1.8(3)$              & $-0.21(5)$            & LO+NLO     & $1$                   & $76.4$                \\
                           & $6.81(4)$             & $-0.749(8)$           & ---                   & NLO+NNLO   & $1$                   & $0.0578$              \\
    \hdashline
    \mr{2}{*}{$5.0$}       & ---                   & $2.33(25)$            & $-0.26(4)$            & LO+NLO     & $1$                   & $63.1$                \\
                           & $8.7(9)$              & $-0.73(15)$           & ---                   & NLO+NNLO   & $1$                   & $31.2$                \\
    \hdashline
    \mr{2}{*}{$6.0$}       & ---                   & $2.99(5)$             & $-0.305(7)$           & LO+NLO     & $1$                   & $1.87$                \\
                           & $10.0(1.9)$           & $-0.6(3)$             & ---                   & NLO+NNLO   & $1$                   & $119$                 \\
    \botrule
\end{tabular*}

%% file: conclusions.tex
\begin{figure}[t]
    \centering
    \begin{subfigure}[t]{0.55\linewidth}
        \centering
        \includegraphics[width=\linewidth]{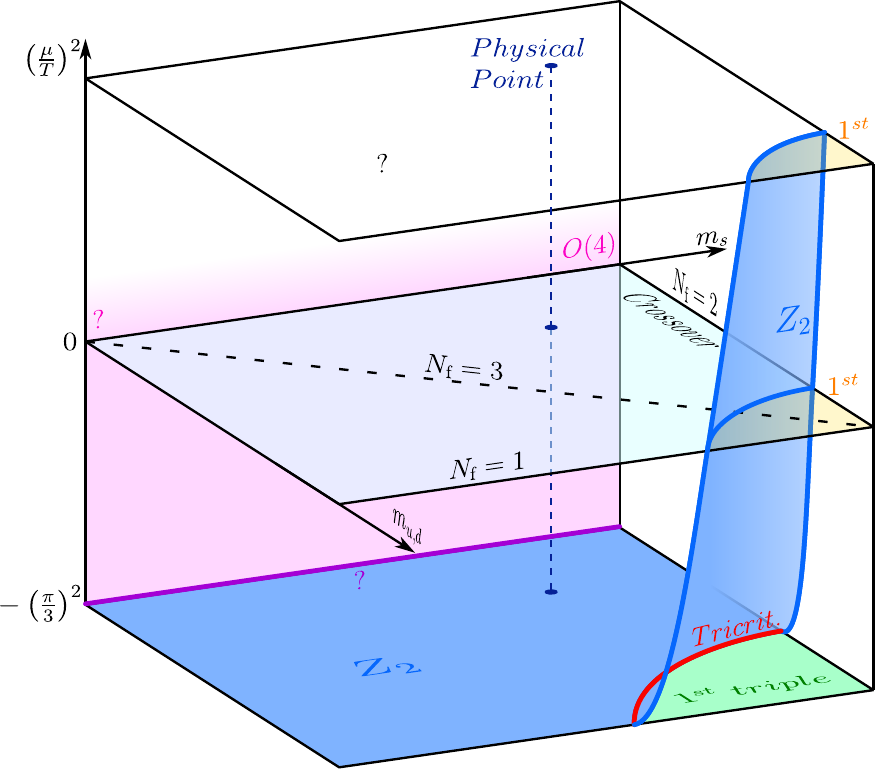}
        \caption{The 3D Columbia plot according to our results.}\label{fig:3dcolumbia_final} 
    \end{subfigure}
    \hfill
    \begin{subfigure}[t]{0.43\linewidth}
        \centering
        \includegraphics[width=\linewidth]{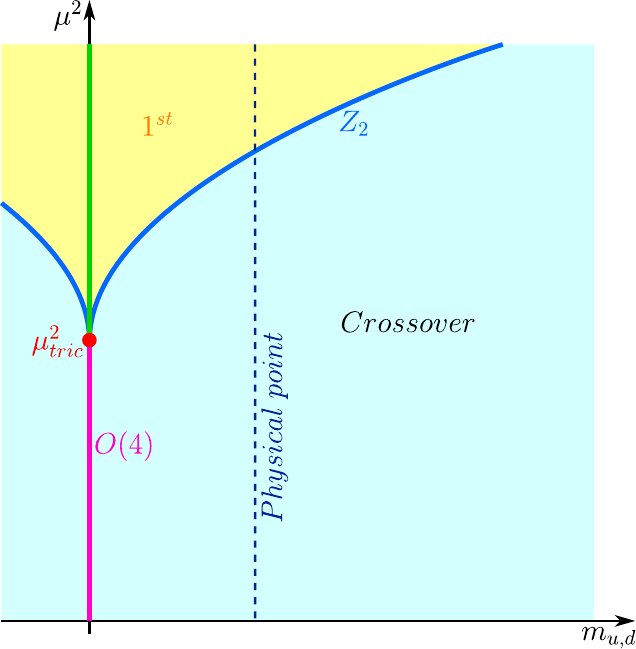}
        \caption{
            Possible scenario for the slice $m_s=m_s^\mathrm{phys}$ allowing 
            for the existence of a critical endpoint in the QCD phase diagram.
        }\label{fig:mu2_v_m}
    \end{subfigure}
    \caption{}
\end{figure}

We have obtained fully non-perturbative information on the nature of the QCD chiral transition as a function of the 
theory's parameters, as predicted by lattice QCD using the standard staggered discretisation on $\nt\in\{4,6,8\}$ lattices.
Both for zero~\cite{Cuteri:2021ikv} and fixed imaginary chemical potential, previously observed first-order transition regions
vanish before the continuum limit is reached.
This identifies them as lattice artefacts, which are to be expected when the chiral limit is approached 
before the continuum limit is taken.
Unless additional (and so far unobserved) first-order transitions will
appear on finer lattices or with chiral fermions, this implies:
\begin{itemize}
    \item The thermal chiral phase transition of $\nf=2+1$ QCD is of second order for $m_{u,d}=0$ and any value of $m_s$,
    turning into an analytic crossover as soon as $m_{u,d}\neq 0$.
    \item The order of the chiral phase transition in $\nf=2+1$ QCD does not show any dependence on 
    imaginary chemical potential, provided our investigated $\mu_i$ value is representative
    for all values up to the Roberge-Weiss point.
\end{itemize} 
This last point is supported by
the absence of non-analytic
chiral transitions in the Roberge-Weiss plane as observed with $\nf=2+1$ improved staggered fermion actions for
$m_\pi\gtrsim 45\unit{MeV}$~\cite{Bonati:2018fvg,Cuteri:2022vwk}, 
and also with 
results from a particular truncation of DSEs in the continuum at real and imaginary
chemical potentials~\cite{Bernhardt:2023hpr,Bernhardt:2025fvk}.
According to the latter works, the chiral critical surface is analytic around
$\mu=0$ and remains in the $m_{u,d}=0$ plane also for small real chemical potentials.
Combining all of this
in a 3D Columbia plot, and barring non-monotonic or oscillatory behaviour of the chiral critical surface, 
the coarse lattice version from \fig\ref{fig:3dcolumbia} changes in 
the continuum limit to the version in \fig\ref{fig:3dcolumbia_final}.

On the other hand, the DSEs in~\cite{Bernhardt:2025fvk} predict the chiral critical surface to bend away from the chiral limit
at large real chemical potentials, in accordance with critical point predictions from functional studies~\cite{Fischer:2018sdj,Gao:2020fbl}.
This scenario requires non-analytic behaviour of the chiral critical surface itself as sketched in \fig\ref{fig:mu2_v_m},
which represents a possible scenario for the slice $m_s=m_s^\mathrm{phys}$ of the 3D~Columbia plot. 
A tricritical point at finite $\mu$ in the chiral limit $m_{u,d}=0$ has indeed been surmised long ago~\cite{Halasz:1998qr}, based on 
analogies with various models for the chiral phase transition.  When the strange quark mass is varied, such a point traces
out a tricritical line. Based on low energy effective models and early lattice results, this line was inferred to be continuously 
connected to the tricritical strange quark mass $m_s^\mathrm{tric}(\mu=0)$
in the putative Columbia plot \fig\ref{fig:columbia_1st}~\cite{Rajagopal:1995bc,Hsu:1998eu,Stephanov:1998dy,Rajagopal:2000wf}. 
Our results, as well as all presently available data from other discretisations and DSE results,  are incompatible with this particular scenario 
and the low energy models predicting it. 

In view of the reduced chiral symmetry of staggered fermions, it would be most valuable
to perform similar studies with chirally symmetric discretisations, as well as to investigate the quark mass dependence
of the critical point observed in Dyson-Schwinger~\cite{Fischer:2018sdj} and functional renormalisation group~\cite{Gao:2020fbl} studies.

%% file: acknowledgments.tex
We acknowledge the help of Alessandro Sciarra to adapt the BaHaMAS 
software~\cite{sciarra_bahamas_2021} to simulations with imaginary chemical potential,
and moving to container-based simulation runs
on the supercomputer.
We also thank the staff of the VIRGO cluster at GSI Darmstadt, where all simulations have been
performed.
We acknowledge use of the analysis software package 
\enquote{Monte Carlo Cpp analysis tools}
by A. Sciarra, C. Pinke, D. Leemueller et al.~\cite{sciarra_2021_17987022} and the
unpublished analysis code \enquote{PLASMA} by 
C. Pinke, F. Cuteri, A. Sciarra et al., as well as the bash utilities repository
\enquote{Script} by A. Sciarra, C. Czaban, F. Cuteri et al.
We thank A. Sciarra for his leading role in maintaining
these packages, and F. Cuteri and A. Sciarra for all contributions to the
groundwork laid in previous projects.
This work is supported by the Deutsche Forschungsgemeinschaft (DFG) through the grant 
CRC-TR 211 \enquote{Strong-interaction matter under extreme conditions} and by the State of Hesse
within the Research Cluster ELEMENTS (Project ID 500/10.006).
Alfredo D'Ambrosio and Reinhold Kaiser acknowledge support by the Helmholtz Graduate School for 
Hadron and Ion Research (HGS-HIRe).

%% file: statistics.tex
\begin{table*}[t]
    \caption{Simulation statistics for $\nt=4$.}\label{tab:nt4-stats}
    \input{nt4-statistsics-table.tex}
\end{table*}

\begin{table*}[t]
    \caption{Simulation statistics for $\nt=6$.}\label{tab:nt6-stats}
    \input{nt6-statistsics-table}
\end{table*}

\begin{table*}[t]
    \caption{Simulation statistics for $\nt=8$.}\label{tab:nt8-stats}
    \input{nt8-statistsics-table}
\end{table*}

In \tabs\ref{tab:nt4-stats},~\ref{tab:nt6-stats} and~\ref{tab:nt8-stats} we provide an overview 
of the simulations and the accumulated statistics for the three simulated values of
$\nt\in\{4,6,8\}$, respectively.

The super-columns report on the different aspect ratios $\ns/\nt$, which we
simulated.
Within these super-columns, each line corresponds to a fixed set of lattice parameters 
$\{\nt,\nf,am,\ns\}$, which encompasses simulations from several $\beta$ values and $4$ independent
Markov chains per $\beta$.
The first entry $\betapc$ is the pseudo-critical gauge coupling determining the phase boundary,
which is obtained from the condition of the vanishing skewness.
Total statistics sums up the number of all generated and analysed trajectories for all $\beta$ 
values of that volume $\ns$.
Next, the number of simulated $\beta$ values is reported.
To check that simulations have been performed on both sides of the phase boundary, the quantity
$|B_3/\sigma_{B_3}|^\mathrm{edge}$ is evaluated on the reweighted data.
It measures the distance of the $B_3$ value at the edge of the simulation range from zero,
expressed in units of the standard deviation of $B_3$.
In the tables we report the minimum value obtained from both edges
$\min(|B_3/\sigma_{B_3}|^\mathrm{edge})$ and we use the values at the peaks of $B_3$, if the
behaviour of $B_3$ is non-monotonic in the simulation range.
To ensure with high confidence that the simulation range includes the transition point, we typically
require $\min(|B_3/\sigma_{B_3}|^\mathrm{edge}) > 3$.
Finally, we calculate the average distance of the $B_4$ values for different Markov chains in units of
the expectation value of the average distance $\bar n_d(B_4)$.
We report the maximum value $\max(\bar n_d(B_4))$ obtained from different $\beta$ values for each
parameter set $\{\nt,\nf,am,\ns\}$.
In the limit of infinite length of the Markov chain, $\bar n_d(B_4)$ approaches $1$.
In practice we demand $\bar n_d(B_4)$ to be not much larger than $2$.
\clearpage

%% file: nt4-statistsics-table.tex
\scriptsize{
{
\setlength{\tabcolsep}{1pt}
\begin{tabular*}{\textwidth}{|l@{\extracolsep{\fill}}l|lllll|lllll|lllll|lllll|}
    \hline
    \mr{2}{*}{$\nf$}    & \mr{2}{*}{$am$}   & $\betapc$ & \mc{5}{l}{Total statistics}       & \mc{4}{l}{\# $\beta$ values}  & \mc{5}{l}{$\min(|B_3/\sigma_{B_3}|^\mathrm{edge})$} & \mc{5}{l|}{$\max \left(\bar n_d(B_4)\right)$} \\\cdashline{3-22}
                        &                   & \mc{5}{c|}{Aspect ratio 3}                    & \mc{5}{c|}{Aspect ratio 4}                    & \mc{5}{c|}{Aspect ratio 5}                    & \mc{5}{c|}{Aspect ratio 6}                \\
    \hline
    \mr{7}{*}{$1.8$}    & 0.0014            & 5.31470   & 520k      & 2 & 1.7 & 1.7       & 5.31495   & 560k      & 2 & 11  & 1.4       &           &           &   &     &           &           &           &   &       &       \\
                        & 0.0017            & 5.31530   & 600k      & 2 & 13  & 1.3       & 5.31545   & 560k      & 2 & 12  & 2.0       & 5.31515   & 600k      & 2 & 8.2 & 1.8       &           &           &   &       &       \\
                        & 0.0020            & 5.31525   & 520k      & 2 & 6.5 & 1.9       & 5.31585   & 732k      & 3 & 14 & 1.5        & 5.31555   & 580k      & 3 & 12 & 1.8        &           &           &   &       &       \\
                        & 0.0023            & 5.31620   & 600k      & 2 & 15  & 0.6       & 5.31600   & 787k      & 3 & 15 & 1.4        & 5.31620   & 480k      & 2 & 9.0 & 1.9       &           &           &   &       &       \\
                        & 0.0026            & 5.31700   & 600k      & 2 & 13  & 0.8       & 5.31650   & 600k      & 2 & 8.6 & 2.6       & 5.31680   & 480k      & 2 & 7.1 & 2.1       & 5.31700   & 467k      & 2 & 6.1 & 2.0     \\
                        & 0.0038            & 5.31945   & 560k      & 3 & 18  & 1.6       & 5.31930   & 440k      & 2 & 2.6 & 1.5       & 5.31885   & 480k      & 2 & 9.8 & 1.4       & 5.31885   & 180k      & 2 & 7.9 & 2.4     \\
                        & 0.0050            & 5.32155   & 520k      & 2 & 10  & 1.4       & 5.32135   & 440k      & 2 & 7.4 & 1.2       & 5.32120   & 540k      & 2 & 12  & 1.9       & 5.32110   & 236k      & 2 & 6.4 & 1.2     \\
    \hline
\end{tabular*}
}
}
\small{
{
\setlength{\tabcolsep}{1pt}
\begin{tabular*}{\textwidth}{|l@{\extracolsep{\fill}}l|lllll|lllll|lllll|}
    \hline
    \mr{2}{*}{$\nf$}    & \mr{2}{*}{$am$}   & $\betapc$ & \mc{4}{l}{Total statistics}       & \mc{2}{l}{\# $\beta$ values}  & \mc{4}{l}{$\min(|B_3/\sigma_{B_3}|^\mathrm{edge})$} & \mc{4}{l|}{$\max \left(\bar n_d(B_4)\right)$} \\\cdashline{3-17}
                        &                   & \mc{5}{c|}{Aspect ratio 2}                    & \mc{5}{c|}{Aspect ratio 3}                    & \mc{5}{c|}{Aspect ratio 4}                    \\
    \hline
    \mr{3}{*}{$1.9$}    & 0.0020            & 5.29495   & 400k      & 2 & 7.0 & 2.0         & 5.29835   & 600k      & 2 & 16 & 1.0          & 5.29850   & 640k      & 3 & 18 & 2.1          \\
                        & 0.0040            & 5.30255   & 400k      & 2 & 12 & 2.6          & 5.30240   & 400k      & 2 & 10 & 2.0          & 5.30205   & 840k      & 4 & 16 & 2.3          \\
                        & 0.0060            & 5.30635   & 400k      & 2 & 12 & 1.3          & 5.30630   & 920k      & 2 & 21 & 3.3          & 5.30550   & 400k      & 2 & 14 & 2.0          \\
    \hline
    \mr{3}{*}{$2.0$}    & 0.0040            & 5.28480   & 400k      & 2 & 14 & 2.4          & 5.28535   & 800k      & 4 & 19 & 1.6          & 5.28535   & 400k      & 2 & 12 & 1.8          \\
                        & 0.0080            & 5.29355   & 400k      & 2 & 11 & 1.9          & 5.29305   & 400k      & 2 & 6.5 & 2.9         & 5.29290   & 800k      & 4 & 13 & 1.2          \\
                        & 0.0120            & 5.30035   & 600k      & 3 & 22 & 1.5          & 5.30000   & 400k      & 2 & 12 & 1.8          & 5.29970   & 800k      & 4 & 19 & 1.4          \\ 
    \hline
    \mr{3}{*}{$2.1$}    & 0.0060            & 5.27190   & 400k      & 2 & 6.7 & 1.2         & 5.27245   & 600k      & 3 & 26 & 2.0          & 5.27205   & 760k      & 3 & 20 & 2.3          \\
                        & 0.0100            & 5.28020   & 400k      & 2 & 12 & 0.84         & 5.28050   & 400k      & 2 & 11 & 1.7          & 5.27975   & 1.0M      & 4 & 15 & 2.1          \\
                        & 0.0130            & 5.28495   & 400k      & 2 & 10 & 1.1          & 5.28545   & 400k      & 2 & 14 & 1.1          & 5.28530   & 400k      & 2 & 12 & 1.7          \\
    \hline
    \mr{3}{*}{$2.2$}    & 0.0080            & 5.26005   & 600k      & 3 & 26 & 1.4          & 5.25870   & 400k      & 2 & 5.8 & 2.2         & 5.25935   & 600k      & 3 & 17 & 1.5          \\
                        & 0.0110            & 5.26615   & 400k      & 2 & 13 & 2.1          & 5.26530   & 400k      & 2 & 7.2 & 2.1         & 5.26495   & 600k      & 3 & 16 & 1.6          \\
                        & 0.0140            & 5.27110   & 400k      & 2 & 14 & 1.8          & 5.27055   & 800k      & 4 & 20 & 2.6          & 5.27055   & 800k      & 4 & 15 & 1.7          \\
    \hline
    \mr{4}{*}{$2.3$}    & 0.0065            & 5.24070   & 800k      & 4 & 19 & 2.1          & 5.24045   & 360k      & 2 & 9.2 & 1.6         &           &           &   &   &               \\
                        & 0.0110            & 5.24985   & 800k      & 4 & 22 & 1.7          & 5.24905   & 400k      & 2 & 6.9 & 1.5         & 5.24900   & 800k      & 2 & 14 & 2.2          \\
                        & 0.0170            & 5.26055   & 800k      & 4 & 25 & 2.3          & 5.26035   & 400k      & 2 & 8.7 & 2.0         & 5.26005   & 800k      & 2 & 15 & 1.4          \\
                        & 0.0230            & 5.27070   & 800k      & 4 & 24 & 1.6          & 5.27055   & 800k      & 2 & 10 & 2.2          & 5.27025   & 800k      & 2 & 14 & 1.7          \\
    \hline
    \mr{4}{*}{$3.6$}    & 0.0370            & 5.10820   & 400k      & 2 & 7.0 & 1.9         & 5.10760   & 800k      & 2 & 11 & 1.6          & 5.10800   & 1.12M     & 3 & 14 & 3.1          \\
                        & 0.0420            & 5.11865   & 400k      & 2 & 5.2 & 1.7         & 5.11800   & 400k      & 2 & 8.2 & 2.0         & 5.11750   & 800k      & 2 & 6.9 & 2.4         \\
                        & 0.0470            & 5.12790   & 400k      & 2 & 8.8 & 1.9         & 5.12765   & 400k      & 2 & 7.9 & 1.2         & 5.12705   & 560k      & 2 & 9.8 & 0.99        \\
                        & 0.0520            & 5.13710   & 400k      & 2 & 11 & 2.1          & 5.13650   & 960k      & 3 & 19 & 1.6          & 5.13645   & 960k      & 3 & 15 & 1.9          \\
    \hline
    \mr{4}{*}{$4.0$}    & 0.0500            & 5.08080   & 400k      & 2 & 4.7 & 1.9         & 5.08075   & 1.2M      & 3 & 20 & 2.2          & 5.08085   & 1.2M      & 4 & 15 & 2.4          \\
                        & 0.0550            & 5.09075   & 400k      & 2 & 12 & 2.7          & 5.09080   & 600k      & 3 & 21 & 1.6          & 5.09060   & 1.4M      & 3 & 25 & 2.1          \\
                        & 0.0600            & 5.10075   & 400k      & 2 & 12 & 1.2          & 5.10020   & 1.0M      & 3 & 23 & 1.3          & 5.10025   & 1.0M      & 3 & 19 & 1.8          \\
                        & 0.0650            & 5.10950   & 400k      & 2 & 8.8 & 2.0         & 5.10935   & 640k      & 2 & 16 & 2.2          & 5.10930   & 600k      & 3 & 17 & 1.7          \\
    \hline
    \mr{3}{*}{$4.5$}    & 0.0650            & 5.04860   & 800k      & 2 & 7.9 & 1.7         & 5.04815   & 400k      & 2 & 8.6 & 2.0         & 5.04820   & 800k      & 2 & 8.0 & 3.8         \\
                        & 0.0700            & 5.05815   & 600k      & 3 & 20 & 2.0          & 5.05800   & 600k      & 3 & 25 & 2.6          & 5.05775   & 1.1M      & 4 & 19 & 2.0          \\
                        & 0.0750            & 5.06735   & 600k      & 3 & 19 & 1.8          & 5.06770   & 600k      & 3 & 16 & 1.8          & 5.06715   & 600k      & 3 & 14 & 1.8          \\
    \hline
    \mr{4}{*}{$5.0$}    & 0.0750            & 5.00900   & 800k      & 4 & 27 & 1.9          & 5.00895   & 400k      & 2 & 13 & 1.8          & 5.00910   & 1.6M      & 4 & 17 & 1.7          \\
                        & 0.0850            & 5.02925   & 800k      & 4 & 22 & 1.8          & 5.02905   & 400k      & 2 & 12 & 1.3          & 5.02875   & 800k      & 2 & 10 & 1.1          \\
                        & 0.0950            & 5.04805   & 800k      & 4 & 27 & 2.4          & 5.04755   & 400k      & 2 & 12 & 1.2          & 5.04770   & 800k      & 2 & 15 & 0.99         \\
                        & 0.1050            & 5.06605   & 400k      & 2 & 8.1 & 1.7         & 5.06575   & 400k      & 2 & 11 & 1.1          & 5.06555   & 800k      & 2 & 11 & 2.2          \\
    \hline
    \mr{3}{*}{$6.0$}    & 0.1000            & 4.94985   & 800k      & 4 & 21 & 2.1          & 4.95010   & 400k      & 2 & 10 & 2.7          & 4.95000   & 1.6M      & 4 & 19 & 1.7          \\
                        & 0.1150            & 4.98020   & 800k      & 2 & 1.6 & 1.0         & 4.98025   & 400k      & 2 & 11 & 1.8          & 4.97980   & 800k      & 2 & 14 & 1.4          \\
                        & 0.1300            & 5.00830   & 400k      & 2 & 7.5 & 1.5         & 5.00785   & 400k      & 2 & 10 & 2.0          & 5.00800   & 800k      & 2 & 9.8 & 2.6         \\
    \hline
\end{tabular*}
}
}

%% file: nt6-statistsics-table.tex
\small{
{
\setlength{\tabcolsep}{1pt}
\begin{tabular*}{\linewidth}{|l@{\extracolsep{\fill}}l|lllll|lllll|lllll|}
    \hline
    \mr{2}{*}{$\nf$}    & \mr{2}{*}{$am$}   & $\betapc$ & \mc{4}{l}{Total statistics}       & \mc{2}{l}{\# $\beta$ values}  & \mc{4}{l}{$\min(|B_3/\sigma_{B_3}|^\mathrm{edge})$} & \mc{4}{l|}{$\max(\bar n_d(B_4))$} \\\cdashline{3-17}
                        &                   & \mc{5}{c|}{Aspect ratio 2}                    & \mc{5}{c|}{Aspect ratio 3}                    & \mc{5}{c|}{Aspect ratio 4}                    \\
    \hline
    \mr{3}{*}{$3.0$}    & 0.0010            & 5.22400   & 360k      & 3 & 13 & 2.7          & 5.22510   & 560k      & 3 & 16 & 1.7          &           &           &   &   &               \\
                        & 0.0020            & 5.22860   & 240k      & 2 & 9.4 & 1.8         & 5.22905   & 364k      & 2 & 6.8 & 2.0         & 5.22850   & 560k      & 3 & 10 & 1.6          \\
                        & 0.0030            & 5.23330   & 240k      & 2 & 7.9 & 1.9         & 5.23290   & 464k      & 2 & 9.9 & 2.1         & 5.23285   & 680k      & 4 & 16 & 1.7          \\
    \hline
    \mr{4}{*}{$3.3$}    & 0.0030            & 5.18145   & 240k      & 2 & 7.7 & 1.9         & 5.18205   & 560k      & 2 & 10 & 1.5          & 5.18195   & 1.04M     & 3 & 11 & 3.7          \\
                        & 0.0045            & 5.18840   & 280k      & 2 & 5.0 & 1.3         & 5.18850   & 1.04M     & 3 & 14 & 2.7          & 5.18845   & 1.0M      & 4 & 14 & 1.4          \\
                        & 0.0060            & 5.19475   & 400k      & 2 & 12 & 1.2          & 5.19520   & 600k      & 3 & 13 & 1.6          & 5.19485   & 660k      & 2 & 8.0 & 1.6         \\
                        & 0.0075            & 5.20060   & 400k      & 2 & 12 & 1.5          & 5.20080   & 400k      & 2 & 4.4 & 1.3         & 5.20090   & 600k      & 3 & 4.9 & 1.7         \\
    \hline
    \mr{3}{*}{$3.6$}    & 0.0050            & 5.14120   & 880k      & 3 & 17 & 1.2          & 5.14125   & 1.2M      & 3 & 12 & 1.9          & 5.14115   & 1.1M      & 3 & 7.4 & 2.0         \\
                        & 0.0075            & 5.15245   & 400k      & 2 & 10 & 1.0          & 5.15205   & 600k      & 3 & 13 & 1.6          & 5.15170   & 880k      & 3 & 14 & 1.9          \\
                        & 0.0100            & 5.16195   & 400k      & 2 & 6.7 & 1.4         & 5.16205   & 1.02M     & 3 & 14 & 1.7          & 5.16195   & 600k      & 3 & 11 & 1.8          \\
    \hline
    \mr{3}{*}{$4.0$}    & 0.0075            & 5.08810   & 400k      & 2 & 6.1 & 1.8         & 5.08810   & 1.06M     & 3 & 11 & 1.6          & 5.08820   & 1.2M      & 3 & 23 & 4.7          \\
                        & 0.0100            & 5.09810   & 400k      & 2 & 9.6 & 3.0         & 5.09850   & 1.0M      & 3 & 24 & 0.87         & 5.09825   & 800k      & 4 & 17 & 1.5          \\
                        & 0.0125            & 5.10865   & 400k      & 2 & 12 & 0.52         & 5.10855   & 800k      & 4 & 18 & 1.9          & 5.10850   & 840k      & 3 & 6.8 & 1.6         \\
    \hline
    \mr{3}{*}{$4.5$}    & 0.0100            & 5.02150   & 560k      & 2 & 2.8 & 1.1         & 5.02195   & 1.04M     & 3 & 20 & 1.5          & 5.02260   & 1.6M      & 5 & 32 & 2.5          \\
                        & 0.0140            & 5.03815   & 840k      & 3 & 20 & 1.8          & 5.03800   & 1.08M     & 3 & 17 & 2.5          & 5.03825   & 1.2M      & 4 & 9.8 & 3.4         \\
                        & 0.0180            & 5.05365   & 400k      & 2 & 5.3 & 2.2         & 5.05375   & 600k      & 3 & 17 & 1.9          & 5.05360   & 1.06M     & 3 & 12 & 2.2          \\
    \hline
    \mr{3}{*}{$5.0$}    & 0.0175            & 4.97865   & 800k      & 4 & 23 & 1.6          & 4.97855   & 1.6M      & 4 & 21 & 3.1          & 4.97850   & 1.2M      & 4 & 18 & 3.6          \\
                        & 0.0210            & 4.99170   & 400k      & 2 & 10 & 1.4          & 4.99230   & 560k      & 2 & 7.4 & 2.1         & 4.99220   & 1.2M      & 4 & 13 & 2.0          \\
                        & 0.0250            & 5.00680   & 1.28M     & 4 & 19 & 2.7          & 5.00755   & 400k      & 2 & 7.3 & 1.2         & 5.00725   & 1.6M      & 4 & 13 & 1.6          \\
    \hline
    \mr{3}{*}{$6.0$}    & 0.0300            & 4.88800   & 400k      & 2 & 9.8 & 1.8         & 4.88838   & 800k      & 4 & 20 & 2.4          & 4.88834   & 1.6M      & 4 & 11 & 4.8          \\
                        & 0.0340            & 4.90350   & 400k      & 2 & 6.7 & 1.2         & 4.90340   & 400k      & 2 & 8.5 & 1.4         & 4.90358   & 1.2M      & 4 & 15 & 3.0          \\
                        & 0.0380            & 4.91790   & 400k      & 2 & 8.5 & 1.4         & 4.91815   & 400k      & 2 & 8.6 & 1.4         & 4.91860   & 960k      & 2 & 3.1 & 0.85        \\
    \hline
\end{tabular*}
}
}

%% file: nt8-statistsics-table.tex
\small{
{
\setlength{\tabcolsep}{1pt}
\begin{tabular*}{\linewidth}{|l@{\extracolsep{\fill}}l|lllll|lllll|lllll|}
    \hline
    \mr{2}{*}{$\nf$}    & \mr{2}{*}{$am$}   & $\betapc$ & \mc{4}{l}{Total statistics}       & \mc{2}{l}{\# $\beta$ values}  & \mc{4}{l}{$\min(|B_3/\sigma_{B_3}|^\mathrm{edge})$} & \mc{4}{l|}{$\max(\bar n_d(B_4))$} \\\cdashline{3-17}
                        &                   & \mc{5}{c|}{Aspect ratio 2}                    & \mc{5}{c|}{Aspect ratio 3}                    & \mc{5}{c|}{Aspect ratio 4}                    \\
    \hline
    \mr{5}{*}{$4.0$}    & 0.0010            & 5.10788   & 180k      & 2 & 8.0 & 1.7         &           &           &   &   &               &           &           &   &   &               \\
                        & 0.0015            & 5.11150   & 200k      & 2 & 8.3 & 1.6         & 5.11100   & 247k      & 2 & 5.9 & 1.2         &           &           &   &   &               \\
                        & 0.0020            & 5.11485   & 360k      & 3 & 12 & 1.8          & 5.11445   & 440k      & 2 & 8.3 & 1.7         &           &           &   &   &               \\
                        & 0.0025            & 5.11790   & 400k      & 2 & 9.2 & 1.9         & 5.11815   & 220k      & 2 & 7.7 & 1.0         & 5.11840   & 319k      & 2 & 3.9 & 2.4         \\
                        & 0.0030            & 5.12190   & 400k      & 2 & 6.4 & 0.9         & 5.12185   & 240k      & 2 & 5.5 & 2.5         & 5.12200   & 380k      & 2 & 7.5 & 2.6         \\
    \hline
    \mr{3}{*}{$4.5$}    & 0.0020            & 5.02845   & 760k      & 4 & 14 & 2.3          & 5.02835   & 540k      & 2 & 9.0 & 1.1         &           &           &   &   &               \\
                        & 0.0030            & 5.03585   & 400k      & 2 & 3.3 & 2.2         & 5.03590   & 675k      & 3 & 11 & 1.1          & 5.03620   & 373k      & 2 & 3.3 & 0.93        \\
                        & 0.0040            & 5.04395   & 320k      & 2 & 3.1 & 1.4         & 5.04355   & 600k      & 3 & 7.5 & 2.8         & 5.04365   & 480k      & 3 & 8.9 & 1.6         \\
    \hline
    \mr{4}{*}{$5.0$}    & 0.0040            & 4.96050   & 600k      & 4 & 15 & 1.9          & 4.96035   & 800k      & 2 & 4.7 & 3.1         &           &           &   &   &               \\
                        & 0.0050            & 4.96755   & 400k      & 2 & 9.8 & 1.3         & 4.96795   & 600k      & 4 & 13 & 3.0          & 4.96775   & 880k      & 4 & 9.9 & 1.3         \\
                        & 0.0060            & 4.97510   & 680k      & 4 & 13 & 2.4          & 4.97525   & 400k      & 2 & 7.2 & 1.6         & 4.97510   & 640k      & 2 & 1.5 & 0.85        \\
                        & 0.0075            & 4.98560   & 560k      & 4 & 12 & 1.6          & 4.98555   & 400k      & 2 & 5.4 & 1.6         & 4.98585   & 400k      & 2 & 3.5 & 1.0         \\
    \hline
    \mr{3}{*}{$5.5$}    & 0.0050            & 4.88695   & 400k      & 2 & 14 & 1.5          & 4.88715   & 1.2M      & 4 & 9.5 & 2.9         &           &           &   &   &               \\
                        & 0.0065            & 4.89785   & 400k      & 2 & 9.3 & 2.5         & 4.89810   & 560k      & 2 & 5.0 & 2.0         & 4.89810   & 560k      & 2 & 8.1 & 2.0         \\
                        & 0.0075            & 4.90490   & 680k      & 3 & 12 & 1.9          & 4.90510   & 400k      & 2 & 5.1 & 2.5         & 4.90525   & 800k      & 2 & 4.0 & 1.2         \\
    \hline
    \mr{3}{*}{$6.0$}    & 0.0060            & 4.81595   & 760k      & 2 & 7.2 & 2.1         & 4.81650   & 1.04M     & 4 & 44 & 3.8          &           &           &   &   &               \\
                        & 0.0080            & 4.83035   & 960k      & 4 & 16 & 2.4          & 4.83044   & 960k      & 4 & 18 & 2.0          & 4.83048   & 800k      & 2 & 7.6 & 2.3         \\
                        & 0.0100            & 4.84400   & 400k      & 2 & 3.6 & 1.2         & 4.84425   & 800k      & 2 & 2.0 & 2.0         & 4.84438   & 560k      & 3 & 7.8 & 1.5         \\
    \hline
\end{tabular*}
}
}